\documentclass[a4paper]{jpconf}
\usepackage{graphicx}

\begin{document}
\title{Quantum models as classical cellular automata}

\author{Hans-Thomas Elze}

\address{Dipartimento di Fisica ``Enrico Fermi'', Universit\`a di Pisa,  
Largo Pontecorvo 3, I-56127 Pisa, Italia}

\ead{elze@df.unipi.it} 

\begin{abstract}
A synopsis is offered of the properties of discrete and integer-valued, hence ``natural'', cellular automata (CA). A particular class comprises the ``Hamiltonian CA'' with discrete updating rules that resemble Hamilton's equations. The resulting dynamics is {\it linear} like the unitary evolution described by the Schr\"odinger equation. Employing Shannon's Sampling Theorem, we construct an invertible {\it map} between such CA and continuous quantum mechanical models which incorporate a fundamental {\it discreteness} scale $l$. Consequently, there is a one-to-one correspondence of quantum mechanical and CA conservation laws. We discuss the important issue of linearity, recalling that nonlinearities imply nonlocal effects in the continuous quantum mechanical description of intrinsically local discrete CA -- requiring locality entails linearity. The admissible CA observables and the existence of solutions of the $l$-dependent dispersion relation for stationary states are 
mentioned, besides the construction of multipartite CA obeying the Superposition Principle. We point out problems when trying to match the deterministic CA here to those envisioned in 't Hooft's {\it CA Interpretation of Quantum Mechanics}.
\end{abstract}

\section{Introduction}
A novel interpretation of quantum mechanics (QM) -- which amounts to redesigning the foundations of quantum theory in accordance with ``classical'' 
concepts, foremost with determinism -- has recently been laid out 
by G.\,'t\,Hooft \cite{tHooft2014}. 
The hope for a comprehensive theory expressed there is founded on the observation that quantum mechanical features arise in a large variety of deterministic ``mechanical'' 
models.  
While practically all of these models have been singular cases, i.e., which cannot easily be 
generalized to cover a realistic range of phenomena incorporating interactions, 
cellular automata (CA) promise to provide the necessary versatility, as we shall discuss 
\cite{PRA2014,EmQM15}. For an incomplete list of various earlier attempts in this field, 
see, for example,  
Refs.\,\cite{H1,H2,H3,Kleinert,Elze,Groessing,Jizba,Wetterich,Mairi,Isidro} and further 
references therein. 

The linearity of quantum mechanics is a fundamental feature most notably 
embodied in the Schr\"odinger equation. This linearity does not depend on the 
particular object  under study, provided it is sufficiently isolated from 
anything else. This is reflected in the Superposition Principle and   
entails the ``quantum essentials''  of interference and entanglement.    

Nevertheless, the linearity of QM has been questioned repeatedly and nonlinear modifications 
have been proposed, {\it e.g.} in Ref.\,\cite{Weinberg,ElzeNonlin}  -- not only as suitable approximations for complicated 
many-body dynamics, but especially in order to test experimentally the  
robustness of QM against such {\it nonlinear deformations},  
as well as in attempts to address the measurement problem in dynamical schemes (none of which has been generally accepted). 
This has been thoroughly discussed by T.F.\,Jordan who presented a proof 
`from within' quantum theory that the theory has to be linear, given the 
essential {\it separability}   
assumption ``... that the system we are considering can be described as
part of a larger system without interaction with the rest of the larger 
system.''\,\cite{Jordan}. 

Recently, we have considered a seemingly unrelated {\it discrete} 
dynamical theory which deviates drastically from quantum theory, 
at first sight. 
However, we have shown with the help of sampling theory 
that the deterministic mechanics of the class of 
Hamiltonian CA can be 
related to QM in the presence of a fundamental time scale. This 
relation demonstrates that consistency of the action principle of the underlying discrete dynamics implies, in particular,  the linearity of both theories.  

Our CA approach  may offer additional insight into  interference and entanglement, in the 
limit when the discreteness scale can be considered as small.  

In Section\,2., we summarize results obtained so far. While in Section\,3., 
we present some considerations about the exact solutions of the CA equations of motion. In particular, we will comment about the relation to ontological CA and their states, the existence of which was introduced as the working hypothesis in Ref.\,\cite{tHooft2014}. 


\section{Natural Hamiltonian CA -- from the action to multipartite automata}
Reasoning about the linearity of QM has led us to the consideration of CA models, which are based on three essential ingredients \cite{PRA2014,EmQM15}: {\bf i}) Deterministic discrete mechanics as proposed by T.D.\,Lee in his program aimed to arrive at a manageable discretization for quantum gravity \cite{Lee,TimeMach} (and references therein); it presumes the existence of a {\it minimal time} scale $l$ and {\it discrete updating rules} for dynamics. {\bf ii}) Sampling Theory \cite{Shannon,Jerri} for discrete structures on or of spacetime \cite{Kempf}; we employ this to construct a {\it map} between CA and QM models in the continuum, which obtain finite-$l$ corrections. {\bf iii}) The ``oscillator representation'' of QM based on decomposing wave functions into real and imaginary parts, $\psi\equiv x+ip$ \cite{Heslot85}; this is suggestive of a relation between apparently classical CA and quantum mechanical models. 

To summarize, the following results hold for a particular class of CA (to be specified): 
\begin{itemize} 
\item Evolution of these CA can be described by a {\it continuous time} Schr\"odinger equation modified by $l$-dependent {\it higher-order derivatives} with respect to time.  
\item There is a $l$-dependent {\it dispersion relation} for stationary states. 
\item There are $l$-dependent {\it conservation laws} in one-to-one correspondence with those of the corresponding QM models in the continuum. 
\item There are {\it multipartite CA} obeying the Superposition Principle, {\it i.e.} which  show the tensorial structures of QM. 
\item If space is discrete as well (assuming the same scale $l$ for simplicity), a generalized Uncertainty Principle can be derived from Robertson's inequality, 
$\Delta A\Delta B\geq | \langle [A,B]\rangle |/2$, with  
$\Delta A:=(\langle A^2\rangle -\langle A\rangle^2)^{1/2}$, {\it etc.}; let $X_{rs}:=lr\delta_{r,s}$ ($r,s$ numbering spatial sites) and $P_{rs}:=-i(\delta_{r,s-1}- \delta_{r,s+1})/2l$, then: 
$\Delta X\Delta P\geq |1+l^2\langle P^2\rangle /2|$, which yields a {\it minimum uncertainty} $\Delta X_{min}=l/\sqrt 2$ \cite{David}. 
\end{itemize} 
Last not least, we find: 
\begin{itemize} 
\item The known QM results are obtained correctly in the {\it continuum limit}, 
$l\rightarrow 0$. 
\end{itemize}  

\subsection{The CA Action Principle} 
We describe natural Hamiltonian CA with countably many degrees 
of freedom presently in terms of {\it complex integer-valued} state variables 
$\psi_n^\alpha$, {\it i.e.} by {\it Gaussian integers} ($\{ z|z\equiv m+in,\; m,n\in {\mathbf Z}\}$), where $\alpha\in {\mathbf N_0}$ denote different degrees of freedom and $n\in {\mathbf Z}$ different states labelled by this discrete {\it clock variable}.  Various equivalent forms of the action for such CA exist, as indicated  earlier \cite{PRA2014}. -- In particular, making use of the mentioned decomposition into real and imaginary parts, $\psi\equiv x+ip$, our theory describes a perfectly classical looking discrete mechanical system, cf. Section\,2.2. However, we will employ here the more compact description in terms of complex variables, which is useful for the  construction of composite CA in analogy with multipartite QM systems.   
   
Let $\hat H:=\{  H^{\alpha\beta}\}$ denote a self-adjoint matrix of Gaussian integers 
that will play the role of Hamilton operator. Furthermore, let us introduce 
$\dot O_n:=O_{n+1}-O_{n-1}$, for any quantity $O_n$ depending 
on the clock variable $n$. With a  summation convention for Greek indices, 
$r^\alpha s^\alpha\equiv\sum_\alpha r^\alpha s^\alpha$, we often  
simplify  notation further by writing 
$\psi_n^{*\alpha}H^{\alpha\beta}\psi_n^\beta\equiv\psi_n^*\hat H\psi_n$. -- 
Then, with $\psi_n^\alpha$ and $\psi_n^{*\alpha}$ as independent variables, 
the CA action ${\cal S}$ is defined by:  
\begin{equation}\label{action} 
{\cal S}[\psi ,\psi^*]\; :=\;
\sum_n\big [\frac{1}{2i}(\psi_n^*\dot\psi_n-\dot\psi_n^*\psi_n)+\psi_n^*\hat H\psi_n\big ] 
\;\equiv\;\psi^*\hat{\cal S}\psi 
\;\;, \end{equation} 
with $\hat{\cal S}$ an useful abbreviation later on. 
In order to set up the variational principle, we introduce    
{\it integer-valued} variations $\delta f$ applied to a polynomial $g$: 
\begin{equation}\label{variation} 
\delta_{f}g(f):=[g(f+\delta f)-g(f-\delta f)]/2\delta f 
\;\; , \end{equation} 
and $\delta_fg\equiv 0$, if $\delta f=0$. -- 
Remarkably, variations  of terms that are 
{\it constant, linear, or quadratic} in integer-valued variables  yield analogous results as  
standard infinitesimal variations of corresponding expressions in the continuum.  
  
With the help of these ingredients, the variational principle is postulated:      
\vskip 0.2cm \noindent 
[{\it CA Action Principle}] \hskip 0.15cm   
The discrete evolution of a CA is determined by stationarity of the  
action under arbitrary integer-valued variations of all 
dynamical variables, $\delta {\cal S}=0$.\hfill $\bullet$ \vskip 0.2cm  

We emphasize the following characteristics of this CA Action Principle:  
\begin{itemize}   
\item While infinitesimal variations do not conform with integer valuedness, 
there is {\it a priori} no restriction of integer variations. Hence {\it arbitrary}  
integer-valued variations are admitted.  
\item One could imagine higher order terms in the action (\ref{action}), {\it i.e.} of 
higher than second order in $\psi_n$ or $\psi_n^*$. 
However, given arbitrary  variations 
$\delta \psi_n^\alpha$ and $\delta \psi_n^{*\alpha}$, 
such additional contributions cannot be admitted. 
Otherwise the number of equations of motion 
generated by variation of the action, according to Eq.\,(\ref{variation}), 
would exceed the number of variables. (A suitable small number of such terms, which are nonzero only for some fixed values of $n$, could encode {\it initial conditions} for the evolution.)    
\end{itemize}  

These features of the CA Action Principle  
are essential in constructing a map between Hamiltonian CA and equivalent quantum mechanical continuum models \cite{PRA2014}. -- For curiosity, generalizations of the 
variations defined in Eq.\,(\ref{variation}) have been considered, in order to allow  
higher than second order polynomial terms in the action. While leading to consistent 
discrete equations of motion, however, these nonlinear equations generally are beset 
with undesirable nonlocal features in the corresponding continuum description 
\cite{Discrete14}.    

\subsection{The equations of motion} 
Applying the CA Action Principle to the action ${\cal S}$ of Eq.\,(\ref{action}) with 
variations $\delta\psi_n^*$ and $\delta\psi_n$, {\it cf.} Eq.\,(\ref{variation}), 
gives discrete analogues of the Schr\"odinger equation and its adjoint, respectively: 
\begin{eqnarray}\label{delpsistar} 
\dot\psi_n&=&\frac{1}{i}\hat H\psi_n 
\;\;, \\ [1ex] \label{delpsi} 
\dot\psi_n^*&=&-\frac{1}{i}(\hat H\psi_n)^* 
\;\;, \end{eqnarray} 
recalling that $\hat H=\hat H^\dagger$ and $\dot\psi_n =\psi_{n+1}-\psi_{n-1}$, {\it etc.} 
-- Note that the action ${\cal S}$ vanishes when 
evaluated for solutions of these finite difference equations. 

Setting $\psi_n^\alpha =:x_n^\alpha +ip_n^\alpha$, with real 
integer-valued variables $x_n^\alpha$ and $p_n^\alpha$, and suitably separating real and imaginary parts of Eqs.\,(\ref{delpsistar})--(\ref{delpsi}), leads to discrete equations that superficially resemble Hamilton's equations for a network of coupled 
classical oscillators:   
\begin{equation}\label{xdotCA} 
\dot x_n^\alpha\;=\;h_S^{\alpha\beta}p_n^\beta +h_A^{\alpha\beta}x_n^\beta  
\;\;,\;  
\;\;\dot p_n^\alpha \;=\;-h_S^{\alpha\beta}x_n^\beta +h_A^{\alpha\beta}p_n^\beta  
\;\;, \end{equation}
where we also split the self-adjoint matrix $\hat H$ 
into real integer-valued symmetric and antisymmetric parts, respectively, 
$H^{\alpha\beta}=:h_S^{\alpha\beta}+ih_A^{\alpha\beta}$. -- 
Notwithstanding their appearance, all finite difference equations here are of second order, necessitating {\it two initial values}, unlike in the continuum \cite{Heslot85,Skinner13}. We will address this point in Section\,3. In any case, the form of equations (\ref{xdotCA}) suggested the name {\it Hamiltonian CA}, which is also justified by the fact that  analogues of Poisson brackets and classical like observables can be introduced here 
\cite{WignerSymp13}. 

\subsection{Conservation laws} 
The time-reversal invariant equations of motion of Section\,2.2.  
give rise to conservation laws which are in {\it one-to-one correspondence} with those of 
the Schr\"odinger equation in the continuum. In fact, the following theorem can be easily verified:  

\vskip 0.2cm \noindent 
[{\it Theorem\,A}] \hskip 0.15cm For any matrix $\hat G$ that commutes with 
$\hat H$, $[\hat G,\hat H]=0$, there 
is a {\it discrete conservation law}: 
\begin{equation}\label{Gconserv} 
 \psi_n^{\ast\alpha}G^{\alpha\beta}\dot\psi_n^\beta +
\dot\psi_n^{\ast\alpha}G^{\alpha\beta}\psi_n^\beta =0 
\;\;. \end{equation}  
For self-adjoint $\hat G$,  defined by Gaussian integers, 
this relation is about real integer quantities.\hfill $\bullet$ \vskip 0.2cm

A rearrangement of Eq.\,(\ref{Gconserv}) yields the corresponding 
conserved quantity $q_{\hat G}$: 
\begin{equation}\label{qG} 
q_{\hat G}:=\psi_n^*\hat G\psi_{n-1}+\psi_{n-1}^*\hat G\psi_n
=\psi_{n+1}^*\hat G\psi_n+\psi_n^*\hat G\psi_{n+1}   
\;\;, \end{equation}  
{\it i.e.} a complex (real  for $\hat G=\hat G^\dagger$) integer-valued two-`time' {\it correlation function} which is invariant 
under a shift $n\rightarrow n+m$, $m\in\mathbf{Z}$. -- In particular, for 
$\hat G:=\hat 1$, the conservation law amounts to a constraint on 
the state variables: 
\begin{equation}\label{normal}  
q_{\hat 1}=2\mbox{Re}\;\psi_n^*\psi_{n-1}
=2\mbox{Re}\;\psi_{n+1}^*\psi_n=\mbox{const}   
\;\;. \end{equation}  
This plays a similar role for discrete CA as the familiar {\it normalization}  
of state vectors in continuum QM; {\it cf.} Eqs.\,(\ref{Qcont1})--(\ref{Qcont2}) below.  --   
We also define the symmetrized conserved quantity: 
\begin{equation}\label{Q} 
\psi_n^*\hat{\cal Q}\psi_n :=\frac{1}{2}\mbox{Re}\;\psi_n^*(\psi_{n+1}+\psi_{n-1})  
\equiv\frac{1}{2}\mbox{Re}\;\psi_n^{*\alpha}(\psi_{n+1}^\alpha +\psi_{n-1}^\alpha )  
\;\;, \end{equation} 
for later use.   

\subsection{Continuum representation}
There exists an one-to-one invertible 
{\it map} between the dynamics of discrete Hamiltonian CA and continuum QM in  
presence of a fundamental time scale $l$ \cite{PRA2014,EmQM15,Discrete14}.  
Due to the finite discreteness scale, the continuous time wave functions are  
{\it bandlimited}, {\it i.e.}, their Fourier transforms have only finite support 
in frequency space, $\omega\in [-\pi /l,\pi /l]$. 
Hence, Sampling Theory allows one to reconstruct continuous time signals, 
wave functions $\psi^\alpha (t)$, from their discrete samples, the CA state variables $\psi_n^\alpha$, and {\it vice versa} \cite{Shannon,Jerri,Kempf}. -- Further aspects  
relating discrete and continuous dynamics, in particular concerning models that do not belong to the class of  Hamiltonian CA, have also been discussed in Ref.\,\cite{tHooft2014}. 

The resulting mapping rules obtained 
through the reconstruction formula of Shannon's Sampling Theorem  \cite{Shannon,Jerri} can be summarized as follows:    
\begin{eqnarray}\label{psit} 
\psi_n^\alpha &\longmapsto &\;\psi^\alpha (t)
\;\;, \\ [1ex] \label{npm1}
\psi_{n\pm 1}^\alpha &\longmapsto &
\;\exp\big [\pm l\frac{\mbox{d}}{\mbox{d}t}\big ]\psi^\alpha (t)
=\psi^\alpha (t\pm l) 
\;\;, \\ [1ex] \label{samp} 
\psi^\alpha (nl)&\longmapsto &\;\psi_n^\alpha 
\;\;, \end{eqnarray} 
keeping in mind that the continuum wave function is bandlimited. -- We remark that kernels that differ from the original $sinc$ function can be used to modify the reconstruction formula, turning the sharp bandlimit into other types of cut-off. However, as long as the reconstruction formular remains a linear map, the form of the equations of motion and of the conservation laws does not change. 

With the help of these results, all CA equations are mapped to their  
continuum versions. For example, corresponding to Eqs.\,(\ref{Gconserv})--(\ref{Q}), there exist analogous conservation laws and conserved quantities, which are  
obtained by applying the mapping rules separately to all wave function factors that appear. 
Thus, for example, the Eq.\,(\ref{Q}) yields the conserved quantity: 
\begin{eqnarray}\label{Qcont1}  
\mbox{const}=\psi_n^*\hat{\cal Q}\psi_n&\longmapsto&\; 
\psi^*(t)\hat{\cal Q}\psi (t)
=\mbox{Re}\;\psi^*(t)\cosh\big [l\frac{\mbox{d}}{\mbox{d}t}\big ]\psi (t)
\\ [1ex] \label{Qcont2} 
&\;&=
\psi^{*\alpha}(t)\psi^\alpha (t)
+\frac{l^2}{2}\mbox{Re}\;\psi^{*\alpha}(t)\frac{\mbox{d}^2}{\mbox{d}t^2}\psi^\alpha (t) 
+\mbox{O}(l^4)   
\;\;, \end{eqnarray} 
with $l$-dependent corrections to the continuum limit ($l\rightarrow 0$), namely the usually conserved normalization $ \psi^{*\alpha}\psi^\alpha =\mbox{const}$\,. 
 
Similarly, the Schr\"odinger equation with finite-$l$ correction terms is 
obtained from Eq.\,(\ref{delpsistar}) \cite{PRA2014}: 
\begin{equation}\label{SchroedCont}
2\sinh (l\partial_t)\psi (t)=-i\hat H\psi (t) 
\;\;. \end{equation} 
This leads, in particular, to the $l$-dependent dispersion relation for stationary states:  
\begin{equation} \label{dispersion} 
lE_\alpha =\arcsin (\textstyle{\frac{l\epsilon_\alpha}{2}})=
{\frac{l\epsilon_\alpha}{2}} [1+({\frac{l\epsilon_\alpha}{2}})^2/6
+\mbox{O}\big ( (l\epsilon_\alpha )^{4}\big )]
\;\;, \end{equation} 
giving their energies $E_\alpha$ in terms of the eigenvalues of the diagonalized {\it dimensionless} Hamiltonian, $\hat H\rightarrow\{l\epsilon_\alpha\}$.  We observe an effect of the band limit, $|E_\alpha |\leq \pi /2l\equiv \omega_{max}/2\;$, as expected. 
Interestingly, by inverting Eq.\,(\ref{dispersion}), we find that the eigenvalues $l\epsilon_\alpha$ must be constrained by $-2\leq l\epsilon_\alpha\leq 2$, for all $\alpha$. A complete classification of such  Hamiltonians represented by {\it symmetric matrices} has been provided recently \cite{McKeeSmyth} --  indeed there are classes of them with infinite numbers of members besides irregular ones . For self-adjoint matrices instead, analogous results are only partially known. 

\subsection{From single to multipartite Hamiltonian CA  \cite{Elze16}}  
So far, we have been concerned with isolated single Hamiltonian CA. Of course, we may ask: Can discrete CA form composite multipartite systems? -- 
Only if the answer is `Yes', the quantum features of CA can possibly cover the Superposition Principle to its fullest extent. 

Therefore, we wonder whether not only the {\it linearity} of the evolution law  but 
also the {\it tensor product structure} of composite wave functions finds its analogue here. These are the fundamental ingredients of the usual continuum theory reflected in 
interference and entanglement of states or observables. Which should be recovered in the continuum limit 
($l\rightarrow 0$) of the CA picture, at least. Furthermore, when the discreteness scale $l$ is finite, the dynamics of composites of CA which do not interact with each other should  lead to {\it no spurious correlations} among them -- the principle of ``no correlations without interactions'', which holds in all of known physics. Naturally, this principle does not rule out the possibility of entangled states, presenting {\it the} prototype of quantum correlations, which may have been formed during interactions among parts of a multipartite system or have been present already from initial conditions for its evolution. 

Let us first discuss in more detail the obstructions encountered when trying to formulate composites of Hamiltonian CA along the lines familiar from QM. While the way to overcome them is described in the next Subsection\,2.5.1. 
 
The want-to-be 
discrete time derivative, $\dot O_n:=O_{n+1}-O_{n-1}$, for any quantity $O_n$ depending on the clock variable $n$, which appears all-over in  
CA equations of motion and conservation laws, does not obey the   
{\it Leibniz rule}: 
\begin{equation}\label{Leibniz} 
\dot {[A_nB_n]}=
\dot A_n\textstyle{\frac{B_{n+1}+B_{n-1}}{2}}+
\textstyle{\frac{A_{n+1}+A_{n-1}}{2}}\dot B_n 
\neq \dot A_nB_n +A_n\dot B_n 
\;\;. \end{equation} 
Similar observations can be expected for other definitions, in particular first-order ones, one might come up with.   

The failure of the Leibniz rule for the 
above `derivative' implies that a multi-CA equation of motion analogous to the single-CA Eq.\,(\ref{delpsistar}): 
\begin{equation}\label{PSIeq} 
\dot\Psi_n=\frac{1}{i}\hat H_0\Psi_n 
\;\;, \end{equation} 
where $\hat H_0$  describes a block-diagonal Hamiltonian in the absence of 
interactions among CA, cannot be decomposed in the usual way. Because of Eq.\,(\ref{Leibniz}), {\it factorization} 
of Eq.\,(\ref{PSIeq}) is hindered on the left-hand side, since unphysical correlations  
will be produced among the components of a factorized wave function, such as   
\begin{equation}\label{PSI} 
\Psi_n^{\alpha\beta\gamma\cdots}=
\psi_n^\alpha\phi_n^\beta\kappa_n^\gamma\cdots 
\;\;, \end{equation}  
or for superpositions of such factorized terms. 
 
Furthermore, applying the mapping rules of Section\,2.4.\,, we find that bilinear terms, $\dot{\psi_n\phi_n}$, for example, do not converge to the correct QM expression, when taking the limit $l\rightarrow 0$. Which should be   
$\partial_t(\psi\phi )=(\partial_t\psi )\phi +\psi \partial_t\phi$, in order to allow     
the decoupling of two subsystems that do not interact.  Even on the right-hand side of Eq.\,(\ref{PSIeq}) we encounter this kind of obstruction.     

The latter is a generic problem of nonlinear terms in the equations 
of motion of discrete CA:  {\it The linear map provided by 
Shannon's Theorem does not commute with the multiplication implied by the 
nonlinearities.}  This follows from the 
explicit reconstruction formula (or any variant thereof that is linear) \cite{PRA2014,Shannon,Jerri,Discrete14}.  

\subsubsection{The many-time formulation} 
The  difficulties just pointed out can be traced to the implicit assumption that components of a multipartite CA are {\it synchronized} to the extent that they share a common clock 
variable $n$. A a radical way out of the impasse 
encountered is to resort to a {\it many-time} formalism \cite{Elze16}. This means giving up synchronization among parts of the composite CA by introducing a set of clock variables, 
$\{ n(1),\;\dots ,\;n(m)\}$, one for each one out of $m$ components.  

It may be surprising to find this in the present nonrelativistic context, since the 
many-time formalism has been introduced by Dirac, Tomonaga, and Schwinger in 
their respective formulations of relativistically covariant many-particle 
QM or quantum field theory, where a global synchronization cannot be 
maintained \cite{Dirac,Tomonaga,Schwinger}.   
  
We replace here the single-CA action of Eq.\,(\ref{action}) by 
an integer-valued multipartite-CA action: 
\begin{equation}\label{maction} 
{\cal S}[\Psi ,\Psi^*] :=\Psi^*\big (
\sum_{k=1}^m\hat{\cal S}_{(k)}\;+\;\hat{\cal I}\big )\Psi
\;\;, \end{equation} 
with $\Psi :=\Psi^{\alpha_1\dots\alpha_m}_{n_1\dots n_m}$ and, correspondingly,  $\Psi^*$ as independent {\it Gaussian integer} variables; 
the self-adjoint operator $\hat{\cal I}$ incorporates interactions between different CA; 
whereas
$\hat{\cal S}_{(k)}$ is as introduced in Eq.\,(\ref{action}), with subscript 
$_{(k)}$ indicating that it concerns {\it exclusively} the pair of indices pertaining to 
the $k$-th single-CA subsystem:   
\begin{equation}\label{Sopk} 
\Psi^*\hat{\cal S}_{(k)}\Psi:=
\sum_{\{ n_k\} }\big[ (\mbox{Im}\;
\Psi_{\dots n_{k}\dots}^{*\dots\alpha_{k}\dots}
\;\dot\Psi_{\dots n_k\dots}^{\dots\alpha_k\dots}
\;+\;\Psi_{\dots n_{k}\dots}^{*\dots\alpha_{k},\dots}\;
H_{(k)}^{\alpha_k\beta_k}\Psi_{\dots n_k\dots}^{\dots\beta_k\dots}
\big ]
\;\;, \end{equation} 
with summation over {\it all} clock variables and over Greek indices appearing twice ; the $\dot{\phantom .}$-operation, however, acts only with 
respect to the explicitly indicated $n_k$, $\dot f(n_k):=f(n_k+1)-f(n_k-1)$, while the 
single-CA Hamiltonian, $\hat H_{(k)}$, requires a matrix multiplication, as before.  

Obviously, we can apply the {\it CA Action Principle} also to the present situation with 
the generalized action of Eq.\,(\ref{maction}). This results in the following discrete equations of motion: 
\begin{equation}\label{mEoM} 
\sum_{k=1}^m\dot\Psi_{\dots n_k\dots}^{\dots\alpha_k\dots}
\;=\;\frac{1}{i}\big (\sum_{k=1}^mH_{(k)}^{\alpha_k\beta_k}
\Psi_{\dots n_k\dots}^{\dots\beta_k\dots} 
\;+\;{\cal I}^{\dots\alpha_k\dots\;\beta_1\dots\beta_m}
\Psi_{\dots n_k\dots}^{\beta_1\dots\beta_m} 
\big )
\;\;, \end{equation}  
together with the adjoint equations; here the interaction $\hat{\cal I}$, 
like $\hat H_{(k)}$, is assumed to be independent of the clock variables and    
the $\dot{\phantom .}$-operation acts only with respect to 
$n_k$ in the $k$-th term on the left-hand side.   

We have verified that this many-time formulation avoids  
the problems of a single-time multi-CA equation, {\it cf.} Eq.\,(\ref{PSIeq}) \cite{Elze16}. -- 
In particular, in the absence of interactions with each other,   
between CA subsystems, $\hat{\cal I}\equiv 0$, {\it no unphysical correlations} are introduced among independent CA subsystems. 

Furthermore,    
{\it continuous multi-time equations} corresponding to Eqs.\,(\ref{mEoM}) are obtained 
by applying the mapping rules given in Section\,2.3. to the discrete equations.   
We find no problem of incompatibility between multiplication according to nonlinear 
terms {\it vs.} linear mapping  according to 
{\it Shannon's Theorem}, since a separate mapping is applied for each one of the 
clock variables. This effectively replaces $n_k\rightarrow t_k,\;k=1,\dots,m$, 
where $t_k$ is a continuous real time variable.  In this way, a {\it modified 
multi-time Schr\"odinger equation} is obtained:  
\begin{equation}\label{mSchroed} 
\sum_{k=1}^m\sinh\big [l\frac{\mbox{d}}{\mbox{d}t_k}\big ] \Psi_{\dots t_k\dots}^{\dots\alpha_k\dots}
\;=\;\frac{1}{i}\big (\sum_{k=1}^mH_{(k)}^{\alpha_k\beta_k}
\Psi_{\dots t_k\dots}^{\dots\beta_k\dots} 
\;+\;{\cal I}^{\dots\alpha_k\dots\;\beta_1\dots\beta_m}
\Psi_{\dots t_k\dots}^{\beta_1\dots\beta_m} 
\big )
\;\;, \end{equation}  
where an overall factor of two from the left-hand side has been absorbed into the matrices  on the right. By construction, here $\Psi$ is bandlimited  
with respect to each variable $t_k$. 

Performing the continuum limit, $l\rightarrow 0$, we arrive at the  
multi-time Schr\"odinger equation (one power of $l^{-1}$ providing the 
physical dimension of $\hat H_{(k)}$ and $\hat {\cal I}$) considered by Dirac and 
Tomonaga \cite{Dirac,Tomonaga}. However, when $l$ is fixed and finite, modifications 
in the form of powers of $l\mbox{d}/\mbox{d}t_k$ arise on its left-hand side, 
similarly as in single CA case before.     

In the present nonrelativistic context, it may be appropriate to identify  
$t_k\equiv t,\;k=1,...,m$, in which case the operator on the left-hand side of 
Eq.\,(\ref{mSchroed}), for $l\rightarrow 0$, can be simply replaced by 
$\mbox{d}/\mbox{d}t$. This results in the usual (single-time) 
{\it many-body Schr\"odinger equation}.  
  
Finally, the study of the {\it conservation laws} of the multipartite CA equations of motion 
can be performed along the lines of Section\,2.3. and analogous results have been 
obtained \cite{Elze16}. 

\subsubsection{On the Superposition Principle in composite CA}   
The equivalent discrete or continuous many-time equations (\ref{mEoM}) and 
(\ref{mSchroed}) are both linear in the CA wave function $\Psi$. Therefore, superpositions 
of solutions of these equations also present solutions and the 
{\it Superposition Principle} does indeed hold for multipartite Hamiltonian 
CA.    

As in the case of single CA, this entails the fact that these discrete systems 
 -- with all variables, parameters, {\it etc.} presented by Gaussian integers -- can 
produce {\it interference} effects as in quantum mechanics. Even more interesting, their 
composites can also show {\it entanglement}, which is deemed an essential feature of QM.    
This follows from the form of the equations of motion, which allow for superpositions 
of factorized states. 

A warning is in order, concerning expressions borrowed from QM which we used freely here, such as ``wave functions'' and ``states''. They usually  
invoke the notion of vectors in a {\it Hilbert space}, which turns into a complex 
projective space upon normalization of the vectors. However, 
as has become obvious in Section\,2.3., 
see Eqs.\,(\ref{normal})--(\ref{Q}), and which can be seen similarly in the multipartite 
case, as 
long as the CA are truly discrete ($l\neq 0$), the normalization (squared) of vectors is not among the conserved quantities, hence not applicable, but is replaced by a conserved 
two-time correlation function. 

Furthermore,  
the space of states presently is {\it not} a 
Hilbert space, since it fails in two respects: the vector-space and completeness 
properties are absent.  Instead, the space of states in the presented CA theory can be classified as a 
{\it pre-Hilbert module over the commutative ring of Gaussian integers}  \cite{Elze16}. 

We conclude here that superpositions of states, interference effects, and entanglement, as in QM, all can be found already on the ``primitive'' level 
of the presently considered natural Hamiltonian CA, discrete single or multipartite systems which are characterized by (complex) integer-valued variables and couplings.    

\section{The formal solution of discrete CA equations of motion} 
We recall that the Schr\"odinger equation, 
$\partial_t\psi(t)=-i\hat H\psi(t)$, can be formally solved by exponentiating, 
$\psi(t)=\exp (-i\hat Ht)\psi (0)$. Here we present the analogous formal solution of the CA equation of motion (\ref{delpsistar}), {\it i.e.}, of the second-order finite difference equation $\dot\psi_n=\psi_{n+1}-\psi_{n-1}=-i\hat H\psi_n$, which   
can be solved by elementary means. We obtain: 
\begin{equation}\label{discsol} 
\psi_n=(2\cos\hat\phi )^{-1}\left (\mbox{e}^{-in\hat\phi}[\mbox{e}^{i\hat\phi}\psi_0+\psi_1]
+(-1)^n\mbox{e}^{in\hat\phi}[\mbox{e}^{-i\hat\phi}\psi_0-\psi_1]\right ) 
\;\;, \end{equation}  
with $2\sin\hat\phi :=\hat H$. From this implicit definition of the operator $\hat\phi$,  we read off once again\,\footnote{{\it Cf.} the remarks following Eq.\,(\ref{dispersion}).} that the eigenvalues of an admissible dimensionless Hamiltonian $\hat H$ must be constrained by $ -2\leq l\epsilon_\alpha\leq 2$, {\it i.e.}, in order to have solutions that neither grow nor decay exponentially in $n$. 

We observe that the general solution of our second-order equation is determined by {\it two} initial values, $\psi_0$ and $\psi_1$, unlike the case of the first-order Schr\"odinger equation. In order to better understand this, we consider the continuum limit, letting $l\approx 0$, but keeping $t:=nl$ fixed and the {\it dimensionful}  
eigenvalues $\epsilon_\alpha$ as well. Thus, we have $2\sin\phi_\alpha =\epsilon_\alpha l\approx 0$, {\it i.e.} $\phi_\alpha\approx 0$. In this limit, the solution (\ref{discsol}) correctly exponentiates, if and only if $\psi_1\equiv\psi_0$. 

Therefore, we have to restrict the set of solutions by allowing only the subset of initial values which satisfy the condition $\psi_1\equiv\psi_0$, in order to assure quantum mechanical behaviour in the continuum limit. 
We speculate about the role of solutions with $\psi_1\neq\psi_0$ shortly. 

Rewriting Eq.\,(\ref{discsol}) symbolically as $\psi_n=\hat T(n+1)\psi_1+\hat T(n)\psi_0$, the evolution fulfills  the composition law   
$\psi_n=\hat T(n-m+1)\psi_{m+1}+\hat T(n-m)\psi_m$, even for general initial conditions. This can be demonstrated by induction and corresponds to the semigroup property of unitary evolution in quantum mechanics. 

\subsection{Are Hamiltonian CA of the ontological kind?}   
In order to address this question, we have to define what is meant by `ontological'.  Here, we refer to the fundamental hypothesis of the {\it CA interpretation of quantum mechanics} proposed and elaborated by G.\,'t\,Hooft \cite{tHooft2014}. Namely, there are so-called {\it ontological states which evolve deterministically}. They constitute the physical reality of nature that we wish to explore.\,\footnote{The essence of this hypothesis is summarized in Figure 6 of the book \cite{tHooft2014} and detailed in the ensuing discussion there, pointing towards astonishing implications.}   

We presently consider a CA characterized by a finite number of degrees of freedom. Its corresponding discrete states may serve to define the basis of a Hilbert space. However, reflecting their ontological character, formal superpositions among these states -- which we do not hesitate to work with in QM -- are not admitted. Superpositions of ontological states are {\it not} ontological, they are not ``out there'' in the Universe.  As a consequence, a finite number of ontological states can only evolve by {\it permutations} among themselves!\,\footnote{At present, we are not concerned with the important relations and distinctions between ontological and either quantum mechanical or classical states, which are specified in Ref.\,\cite{tHooft2014}.}  

The question in the title of this section can now be asked more precisely: Is the Hamiltonian CA dynamics, as described, compatible with the permutation dynamics of ontological states? We will find some partial answers by resorting to very simple indicative examples.     

Let $\psi_n^\alpha ,\;\alpha =1,2$ denote the state variables and $\hat H$ a 2x2 self-adjoint Hamiltonian matrix, all (complex) integer-valued, as before in Section\,2. 
It is convenient to collect the state variables into a two-dimensional vector and write the evolution equation (\ref{delpsistar}) as: 
\begin{equation}\label{update} 
\psi_n=\psi_{n-2}-i\hat H\psi_{n-1}\;\;,\;\;\;
\psi_n\equiv \Big (\begin{array}{c}\psi_n^1 \\ \psi_n^2\end{array}\Big ) 
\;\;. \end{equation} 

We have argued that the formal solution of such an equation, given in Eq.\,(\ref{discsol}), approaches an exponential as in QM for the continuum limit  
($l\rightarrow 0$), if and only if the necessary two inital values coincide.  
Assuming this, {\it e.g.} $\psi_1=\psi_0:=(1,0)^t$,  we find that the evolution 
generally takes place in the two-dimensional space given; however, it cannot be avoided that in the sequence $\{\psi_n\}$ appear regularly states 
which are superpositions of any two linearly independent states, say $\psi_m$ and $\psi_{m'}$, which we may select as ontological basis states, no matter how we choose 
$\hat H$. In this case, evolution does {\it not} consist of  
permutations only.  

In other words, in this two-dimensional example, we learn that the discrete evolution which  conforms with QM in the continuum limit passes unavoidably through {superposition states} as intermediates. Which are not ontological. We leave it as a {\it conjecture}  that this happens independently of the dimensionality of the state space. Which seems to disqualify such a Hamiltonian CA  as an example of a strictly ontological model. 

It is tempting to speculate now that an ontological model must deviate in one way or another from the Hamiltonian CA under consideration and from a QM model in its continuum limit. -- One way could be to give up self-adjointness of the Hamiltonian, which we do not follow here. -- Another is to abandon the condition of coinciding initial values, $\psi_1\equiv\psi_0$, thereby renouncing the requirement that such a model have an obvious QM continuum limit, {\it cf.} Section\,2.4.   

For example, we may choose one of the Pauli matrices as Hamiltonian: 
\begin{equation}\label{Hsadj} 
\hat H:=\Big (\begin{array}{c  c} 0 & 1 \\ 1 & 0 \end{array}\Big ) 
\;\;, \end{equation} 
with $\hat H^2=\mathbf{1}$. Then, considering the initial values $\psi_0=(1,0)^t$ and $\psi_1=(0,1)^t$, we simply evaluate the first few steps of the evolution according to Eq.\,(\ref{update}) to obtain: 
\begin{equation}\label{ontolevol}  
\psi_2=(1-i)\psi_0\;,\;\psi_3=-i\psi_1\;,\;\psi_4=-i\psi_0\;,\;\psi_5=-(1+i)\psi_1\;,\;
\psi_6=-\psi_0\;,\;\psi_7=-\psi_1\;,\;\dots 
\;\;. \end{equation}   
It is obvious how the sequence continues and arrives at the initial values within the next 6 steps, thus becoming {\it periodic}. We could have used the general solution (\ref{discsol}) instead, with $\hat\phi =\frac{\pi}{6}\hat H$, in the present example. -- The rescaling in the intermediate states $\psi_2$ and $\psi_5$ would violate the conserved normalization of a state vector in QM. However, it is perfectly in accordance with the corresponding discrete conservation laws discussed in Section\,2.3. and applicable here. -- In this way, we have found an example of an elementary permutation dynamics realized by a Hamiltonian CA. 

It will be interesting to see whether or how the findings here generalize to more complex higher dimensional ontological models.

\section{Conclusion} 
This presents a brief review of earlier work which has demonstrated surprising quantum features arising in integer-valued, hence ``natural'', {\it Hamiltonian cellular 
automata} \cite{PRA2014,EmQM15,Discrete14,WignerSymp13,Elze16}.  

The study of this 
particular class of CA is motivated by 't\,Hooft's {\it CA  
interpretation of quantum mechanics} \cite{tHooft2014} and various recent attempts to construct models 
which may eventually lead to demonstrating that the essential features of QM can 
all be understood to emerge from pre-quantum deterministic dynamics and that its puzzles, such as the measurement problem, can be satisfactorily resolved after all.    

The single CA we have considered allow practically for the first time to 
reconstruct quantum mechanical models  with nontrivial Hamiltonians in terms of 
such systems with a {\it finite discreteness scale}. -- 
Furthermore,  we have extended 
this study by describing {\it multipartite systems}, analogous to many-body QM. 
Not only is this useful for the construction of more complex models {\it per se} 
(especially with 
a richer structure of energy spectra), but it is also necessary, in order to  
extend the Superposition Principle of QM to a description at the CA level. 
We find that it can be introduced already there to the fullest extent, 
compatible with a tensor product structure of multipartite states, 
which entails the possibilities of their {\it interference} and {\it entanglement}. 

Surprisingly, we have been forced -- in our approach employing Sampling Theory to construct the 
map between CA and an equivalent continuum picture -- to introduce a 
many-time formulation, which only appeared in relativistic quantum mechanics before,  
as introduced by Dirac, Tomonaga, and Schwinger \cite{Dirac,Tomonaga,Schwinger}.  
This points towards a crucial further step in these developments, which is still missing, 
namely a relativistic CA model of {\it interacting quantum fields}. Without the possibility of  
interacting multipartite CA with quantumlike features, as described here, it is hard to 
envisage a CA picture of dynamical fields spread out in spacetime.   

Last not least, we have presented a primitive attempt to see the relation, if any, between the Hamiltonian CA here and ontological models as advocated in Ref.\,\cite{tHooft2014}.  
Which resulted in the suggestion that it might depend on initial conditions, whether 
such a CA behaves either as a QM model (with corrections due to discreteness) or is of the ontological kind.  

\ack
It is a pleasure to thank Jack Ng for a discussion, the organizers of the 10th Biennial Conference on Classical and Quantum Relativistic Dynamics of Particles and Fields IARD 2016 (Ljubljana, June 2016) for the invitation to this very nice meeting,  and Martin Land and Matej Pavsic for their kind hospitality.  

 
\end{document}